\documentclass[final,5p,times,twocolumn]{elsarticle}

\usepackage{graphicx}

\usepackage{amssymb}
\usepackage{amsthm}

\usepackage{amsmath}

\journal{Chaos, Soliton Fractals}

\usepackage{bm}
\usepackage{here}
\usepackage{color}

\newcommand{\blue}[1]{\textcolor{blue}{#1}}
\definecolor{mygreen}{rgb}{0.2,0.8,0.2}

\begin{document}

\begin{frontmatter}


\title{Localization in One-Dimensional Tight-Binding Model with Chaotic Binary Sequences
}




 \author{Hiroaki S. Yamada}
\address{Yamada Physics Research Laboratory, Aoyama 5-7-14-205, Niigata 950-2002, Japan}

\begin{abstract}
We have numerically investigated localization properties in the one-dimensional 
tight-binding model with chaotic binary on-site energy sequences 
generated by a modified Bernoulli map with the stationary-nonstationary
chaotic transition (SNCT).
The energy sequences in question might be characterized by their correlation parameter $B$ 
and the potential strength $W$.
The quantum states resulting from such sequences have been characterized in the two ways:
Lyapunov exponent at band centre and the dynamics of  the initially localized wavepacket. 
Specifically, the $B-$dependence of the relevant Lyapunov exponent's decay 
is changing from linear to exponential one  around the SNCT ($B \simeq 2$).
Moreover,  here we show that  even in the nonstationary regime, 
mean square displacement (MSD) of the wavepacket  
is noticeably suppressed in the long-time limit (dynamical localization).
The $B-$dependence of  the dynamical localization lengths determined by 
the MSD exhibits a clear change in the functional behaviour  around SNCT, and
its rapid increase gets much more moderate one for $B \geq 2$.
Moreover we show that the localization dynamics for $B>3/2$ deviates from the 
one-parameter scaling of the localization in the transient region.  
\end{abstract}


\end{frontmatter}


\def\ni{\noindent}
\def\nn{\nonumber}
\def\bH{\begin{Huge}}
\def\eH{\end{Huge}}
\def\bL{\begin{Large}}
\def\eL{\end{Large}}
\def\bl{\begin{large}}
\def\el{\end{large}}
\def\beq{\begin{eqnarray}}
\def\eeq{\end{eqnarray}}

\def\eps{\epsilon}
\def\th{\theta}
\def\del{\delta}
\def\omg{\omega}

\def\e{{\rm e}}
\def\exp{{\rm exp}}
\def\arg{{\rm arg}}
\def\Im{{\rm Im}}
\def\Re{{\rm Re}}

\def\sup{\supset}
\def\sub{\subset}
\def\a{\cap}
\def\u{\cup}
\def\bks{\backslash}

\def\ovl{\overline}
\def\unl{\underline}

\def\rar{\rightarrow}
\def\Rar{\Rightarrow}
\def\lar{\leftarrow}
\def\Lar{\Leftarrow}
\def\bar{\leftrightarrow}
\def\Bar{\Leftrightarrow}

\def\pr{\partial}

\def\Bstar{\bL $\star$ \eL}
\def\etath{\eta_{th}}
\def\irrev{{\mathcal R}}
\def\e{{\rm e}}
\def\noise{n}
\def\hatp{\hat{p}}
\def\hatq{\hat{q}}
\def\hatU{\hat{U}}

\def\iset{\mathcal{I}}
\def\fset{\mathcal{F}}
\def\pr{\partial}
\def\traj{\ell}
\def\eps{\epsilon}




\section{Introduction}
It has been known that in one-dimensional 
disordered systems (1DDS) with uncorrelated on-site disorder 
all eigenstates are exponentially localized 
\cite{ishii73,abrahams79,lifshiz88}.
Still, for the 1D tight-binding model 
 with  potential sequences  generated  by Fourier filtering method (FFM)
it has been known that  correlations arising in the on-site potential  
 delocalize the eigenstates and induce localization-delocalization transition (LDT) 
\cite{moura98,izrailev99,zhang02,shima04,kaya07,kaya09,gong10,iomin09,izrailev12,croy11,deng12,gong12,albrecht12}.
Indeed, the potential sequences involved ought to have long-range correlation with power spectrum $S(f) \sim 1/f^\alpha$ 
(\blue{$f<<1$}, $\alpha \geq 2$), where $f$ denotes frequency
 and $\alpha$ is spectrum index.
The potential sequence is non-stationary  when the total power 
$\int_0^\infty S(f) df$ is divergent. 

Noteworthy, the results do not contradict  the Kotani theory of the localization  
stating that if the stationary random potential is non-deterministic, 
absolutely continuous spectrum is absent. 
The stationarity is a sufficient condition for 
the absence of absolutely continuous spectrum \cite{kotani84}.
On the other hand,  
 the potential sequence characterized by the power spectrum 
with the exponent $\alpha>1$ would be nonstationary.
Further, most recent numerical studies show that the sequences 
with the power-law spectrum generated by Weierstrass function 
with fractal dimension $1<D<2$ induce the LDT 
\cite{garcia09,garcia10,petersen12,petersen13,yamada15}.  

There are systematic numerical studies 
for the above-mentioned 1DDS models with a potential to take continuous value
 like in the Anderson model.
Whereas uncorrelated random model  (e.g., Bernoulli Anderson model) 
with discretized values are well-known to show specific localization phenomena, 
the number of studies of localization and delocalization with 
the correlated binary potential are still few
\cite{yamada91,yamada04,yamada92,oliveira01,sales12,cheraghchi05,esmailpour07,lazoa10}.
E.g., among the latter examples the following one should be mentioned.
There is a study of  delocalization in binary  "0" and  "1" system and 
the sparse potential which takes 
 different values for prime sites only \cite{oliveira01}.
It is in such a case that the very "sparse" model should also naturally 
be "nonstationary".
Remarkably,  the existence of the LDT due to the potential intensity has also been 
demonstrated in the sparse impurity distance model \cite{stollmann01}.

Furthermore, a number of works also have been published on 
 localized and delocalized phenomena in 
1DDS with  deterministic correlated sequence  generated by chaotic map
\cite{yamada91,nazareno98,sakurada01,pinto05,costa11}.
In our earlier papers, we also numerically investigated the localization and delocalization
 phenomena of binary random systems with long-range correlation
by the modified Bernoulli map with stationary-nonstationary chaotic transition (SNCT)
\cite{yamada91}. We shall refer to such a system MB system 
in the following \cite{aizawa84,yamada04,akimoto05}.
The sequence becomes asymptotic non-stationary chaos 
for $\alpha >1$.
In the MB system, it is possible to create the potential sequence that
changes  the property from short-range correlations 
including $\delta-$correlations to long-range correlation 
with a gentle change of the correlation parameter $B$.
Meanwhile, studying in detail the modalities of the localization events,
especially in the situations, where transitions from 
stationary ($3/2<B<2$) to nonstationary regime ($B>2$)  regimen 
takes place in the binary correlated 1DDS has  still not been enough.
In particular, wavepacket dynamics in the nonstationary potential  
has hardly been investigated \cite{santos06,wells08,gong11a,gong11b,gholami17}.
In this paper, we use the long-range correlated MB system 
having the binary potential sequence with taking either one of  
$-W$ or $W$,  like in our previous papers.
We aim at reporting the characteristic $B/W-$dependences of 
the Lyapunov exponent, the normalized localization length (NLL), and 
the quantum diffusion of the initially localized wavepacket around SNCT
in binary correlated disordered systems.


This paper is organized as follows.
In the next section, we shall briefly introduce the modified Bernoulli model.
In Sect.\ref{sec:NLL} we report about the 
global behaviour of the $B-$dependence and $N-$dependence
of  Lyapunov exponent and the NLL at band centre by the numerical calculation.
While the Lyapunov exponent is positive 
throughout all the $B$ regions studied here,  
 the Lyapunov exponent decreases linearly for $ B <2 $, but 
decays exponentially for $B> 2$.
As a result, the quantum states get delocalized ($\gamma_N \to 0$) 
with $B \to \infty$. 
In Sect.\ref{sec:DL},  we report on the dynamical localization phenomena
 in the system.
We find that the MSD is finite and dynamically localized in $t \to \infty$
even if the correlation parameter changes 
from stationary regime $B<2$ with power-law decay of the correlation 
to nonstationary regime ($B \geqq 2$).
Its dynamical localization length (DLL) increases with the correlation parameter $B$, 
but the $B-$dependence changes from a relatively rapid increase 
to a more moderate one around SNCT($B \simeq 2$).
The one-parameter scaling based on the localization length 
has large fluctuation in the transient region  from ballistic motion to localization 
for $ B> 3/2$.
The summary and discussion  are presented in the last section.
Appendix shows the sample fluctuation including nonstationary regime.

\section{Model}
\label{sec:model}
We consider the one-dimensional tight-binding Hamiltonian
 describing single-particle electronic states  as
\begin{eqnarray}
  H= \sum_{n=1}^N Wv(n)c_n^{\dagger}c_n +  \sum_{n=1}^{N-1}c_{n}^{\dagger}c_{n+1} + H.C., 
\label{eq:tight-binding}
\end{eqnarray}
where $c_n^{\dagger}$($c_n$) is the creation (annihilation) operator for 
an electron at site $n$. 
The $\{ v_n \}_{n=0}^{N}$  and $W$ are the disordered on-site energy sequence and 
the  strength, respectively. 
The amplitude of the quantum state $|\Phi>$ is given by 
$\phi(n) \equiv <\Phi|c_n^{\dagger}c_n|\Phi>$ in the site representation.
To model the correlated disorder potential 
for $v_n$($n\leq N$) in Eq.(\ref{eq:tight-binding}), 
we use the modified Bernoulli map;
\beq
& & X_{n+1} =   \nonumber \\
& &
\begin{cases}
 X_{n} + 2^{B-1}(1-2b)X_{n}^{B}+b  & (0 \le X_{n} < 1/2)  \\
   X_{n} - 2^{B-1}(1-2b)(1-X_{n})^{B} -b & (1/2 \le X_{n} \le 1),    
  \end{cases}
\label{eq:map}
\eeq
\noindent
where $B$ is a bifurcation parameter which controls the correlation
of the sequence. $b$ stands for the small perturbation which
is set $b=10^{-13}$ in this paper.
The map has been introduced to investigate the basic property of the
intermittent chaos by Aizawa \cite{aizawa84}.


\begin{figure}[htbp]
\begin{center}
\includegraphics[width=7.5cm]{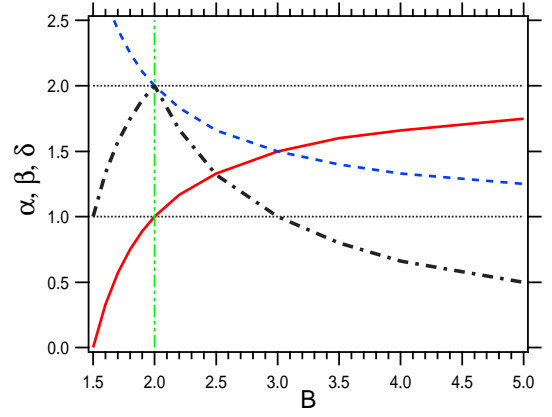}
\caption{
(Color online)
Spectral index $\alpha$ (red solid line),
index $\beta$ (blue broken line) of the residence time distribution,
and $\delta$ (black dotted-dashed line) of the variance of the renewal process 
 as a
function of correlation parameter $B$.
The black dotted lines denote $\alpha=2(=\alpha_c)$ and $\beta=1(=\beta_c)$
in a limit $B \to \infty$.
The vertical double dotted-dashed (green) line denotes the SNCT $B=2$. 
In the present paper localization phenomena in a range $B \in [1.5,3.5]$ is
mainly investigated.
}
\label{fig:alpha-B}
\end{center}
\end{figure}

The sequence is stationary for $B< 2$ and nonstationary for $B\geq 2$. 
The stationary property is recovered by the perturbation though the essential 
property remains invariant for a long time 
$n < n_{b}$, where $n_b \simeq (2b)^{(1-B)/B}$
\cite{aizawa84}.
We use the course-grained binary sequence $\{ v_{n} \}$
 by the following rule:
\beq
 \begin{cases}
 0 \le X_{n}< 1/2 & \to v_{n} = - 1 \\
 1/2 \le X_{n} < 1 & \to v_{n} =  1.
 \end{cases}
\label{eq:binary}
\eeq
\noindent
Accordingly, the statistical property of the binary sequence 
can be characterized by changing the correlation parameter $B$.
The following properties, for example,  are analytically and numerically derived.
In the stationary regime ($3/2<B<2$) 
the correlation function 
of the symbolic sequence decreases obeying the inverse-power law 
with the long-range correlation 
for large $n$ \cite{aizawa84},
\begin{eqnarray}
C(n) 
\equiv <v_{n_0+n}v_{n_0}>
 \sim n^{-\frac{2-B}{B-1}}(n>>1).
\end{eqnarray}
\noindent
The correlation shows the critical decay $C(n) \sim 1/n$ at $B=3/2$.
In the nonstationary regime ($B\geq 2$) the correlation decays
as,
\begin{eqnarray}
 C(n) \simeq 1-\frac{2}{B}(\frac{n}{n_b})^{\frac{B-2}{B-1}}, 
\end{eqnarray}
\noindent
for $n \leq n_b$. 
 The power spectrum
$S(f)=\frac{1}{N}\left|\sum_{n=0}^N e^{-i2\pi f n/N}\right|^2$ ($f=0,1,2,...,N-1$) 
in the low frequency limit
behaves 
\beq
S(f) \sim
 \begin{cases}
 f^0  & 1 \le B < 3/2  \\
   f^{-\alpha}  & 3/2 \le B \le \infty, 
 \end{cases}
\label{eq:power}
\eeq
in the thermodynamic limit ($N \to \infty$),  where 
\beq
 \alpha \simeq \frac{2B-3}{B-1}.
\label{eq:wei-pot}
\eeq
That is, the stationary sequence changes to nonstationary one with $S(f) \sim 1/f$
around $B \simeq 2$.
It is suggested that in FFM model and Weierstrass model with long-range correlation
LDT appear in a case with $\alpha \simeq 2$.
Note that if $B\to \infty$, then $S(f)\sim f^{-2}$ 
as shown in Fig.\ref{fig:alpha-B}.
Still,  the localization property of 1DDS around $\alpha \simeq 1$ 
have not yet been studied.   
In the present paper, we investigate the change of the quantum states 
around the SNCT of the sequenece.
It has already been reported that 
this property of the sequence strongly affects the statistical nature 
of the Lyapunov exponents of the electronic wave functions \cite{yamada91}.

Moreover, 
the binary sequence $\{ v_n \}$ can be recast as 
$\{(m_0,\sigma),(m_1,-\sigma),(m_2,\sigma),(m_3,-\sigma).....\}$.
Here $(m_k,\sigma)$ stands for the $m_k$ times iterating 
of one and the same symbol $\sigma$, where $\sigma$ represents $-1$ or $1$.
The sequence is uniquely determined by the cluster size 
distribution $P(m)$ for the number $m$ of iterations in the pure
sequence $(m,\sigma)$, which is independent of the value of the symbol.
Hence, the time interval $m$ between successive renewal events
is a random variable, whose probability 
density function $P(m)$:
\beq
P(m) \sim m^{-\beta},
\eeq
where 
\beq
\beta=\frac{B}{B-1}.
\eeq
When $B < 2$ the cluster size is finite ($<m> < \infty$), 
and when $B \geq  2$ it diverges.
 It is worth noting that in the stationary regime ($B < 2$) the normalized stationary 
distribution (invariant measure) exists; on the other hand, 
when $B \ge 2$ the sequence becomes nonstationary and the normalizable measure
does not exist when $b=0$.
If the perturbation $b$ is not introduced, the sequence is constructed
by only the pure cluster of the same type symbol with probability one
 for the nonstationary regime.


The number of the renewal events $N_t$ in the interval $[0,t]$
can be approximated by renewal process.
Then the  variance $Var(N_t)=<(N_t-<N_t>)^2>$($\sim t^\delta$)
behaves as $\delta=4-\beta(3/2<B<2)$, $\delta=2\beta-2$($B \geqq 2$) 
depending on the parameter $B$ \cite{tanaka95,akimoto05}.
(See Fig.1 for the $B-$dependence of the exponent $\delta$.)
Indeed, it has been shown numerically that 
$B-$dependence of the exponent $\delta$ becomes maximum at $B=2$ 
and then decreases. 
We examine the localization property 
around the transition point ($B \simeq 2$) in potential sequence.


\section{Lyapunov exponent and normalized localization length}
\label{sec:NLL}
The ensemble-averaged finite size Lyapunov exponent  is defined by 
\begin{eqnarray}
\gamma_N &=&  \left<\frac{ \ln \left( |\phi(N)|^2 + |\phi(N+1)|^2 \right)  }{2N} \right>,
\label{eq:gamma_n}
\end{eqnarray}
for  $N>>1$, where
$\left<...\right>$ denotes the ensemble average over 
uniformly distributed initial value $X_0 \in [0,1]$ in Eq.(\ref{eq:map}).
We obtained the finite size Lyapunov exponent by standard transfer matrix
 products with the initial state $\phi(0)=\phi(1)=1$ \cite{crisanti93}.
Then $\xi(N)(=\gamma_N^{-1})$ denotes 
the finite size localization length (LL).  
We define the NLL to characterize the tail 
of the wavefunction,  
\beq
\Lambda_N \equiv \frac{\xi(N)}{N}.
\eeq
It is useful to study the localization and delocalization property 
that $\Lambda_N$ decreases (increases) with the system size 
$N$ for localized (extended) states,
and it becomes constant for the critical states.
In what follows, we investigate the NLL 
 by changing the system size $N$ and the correlation parameter $B$
for the band centre $E=0$.
The typical size $N$ and ensemble size used 
here are $N=2^{16} \sim 2^{23}$ and $2^{10}\sim 2^{12}$, respectively.
The robustness of the numerical 
calculations has been confirmed in each case.

\begin{figure}[htbp]
\begin{center}
\includegraphics[width=7.0cm]{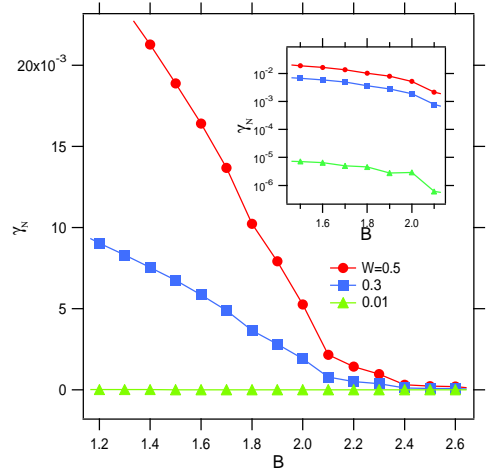}
\caption{(Color online)
Lyapunov exponent $\gamma_N$ in the band centre ($E=0$) as a function of 
the correlation parameter $B$ for $W=0.5, 0.3, 0.01$.
$N=2^{23}$ and the sample size is $2^{12}$. 
The inset of the panel  is semi-log plot.
}
\label{fig:gamma-B-dep}
\end{center}
\end{figure}

\begin{figure}[htbp]
\begin{center}
\includegraphics[scale=0.6]{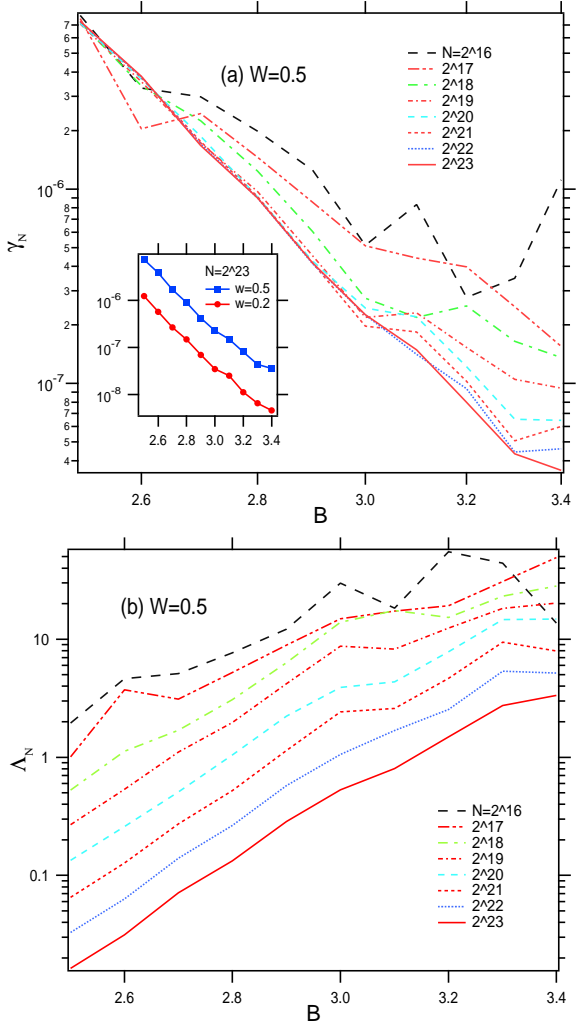}
\caption{(Color online)
(a)Lyapunov exponent $\gamma_N$ and 
(b)the normalized localization length $\Lambda_N$ as a function of 
the correlation parameter $B$ for several system size  $N=2^{16}-2^{23}$.
$W=0.5$, $E=0$, .and the sample size is $2^{12}$. 
Note that the vertical axis is in logarithmic plot.
The inset of the panel (a) is result for $W=0.5$ and $W=0.2$ 
for   $N=2^{23}$. 
The slope is about $-6.38$ in the semi-logarithmic plot.
}
\label{fig:E0-B-dep}
\end{center}
\end{figure}

\subsection{A transition at $B \simeq 2$}
As shown in Fig.\ref{fig:gamma-B-dep},  for $1.2<B<2$ 
the Lyapunov exponent $\gamma_N $ monotonically 
decreases toward zero around $B=2$, and the $B-$dependence 
is roughly estimated as 
\beq
\gamma_N \simeq \gamma_0(W)-k(W)B,
\eeq
where $\gamma_0(W)$ and  $k(W)$ are $W-$dependent coefficients.
The  tendency is almost independent of the potential strength $W$.
In what following we show the more details 
numerical results for the nonstationary regime $B \geqq 2$.


Figure \ref{fig:E0-B-dep}(a) shows the $B-$dependence of the Lyapunov exponent
for the different system size with a fixed value $W=0.5$.
As the system size grows, the $B-$dependence of $\gamma_N$ for  $B \geq 2$ shows 
stable exponential decay such as:
\beq
  \gamma_N(W,B) \simeq \gamma_c(W) \exp\{-cB\},
\eeq
where $\gamma_c(W)$ is the coefficient of dependence on $W$,
 and  $c=6.38$ numerically.
Moreover, from the comparison with the case of  $W=0.2$ shown in the inset,
it turns out that the decay rate  (slope) hardly depends on 
the potential intensity $W$.
That is, $B \to \infty$ is expected to be $\gamma_N \sim 0$ 
corresponding to the delocalized state.

%

Therefore, it seems that  with the increase of $B$ 
the Lyapunov exponent shifts from a linear-decay
 to an exponential-decay in the MB 
 quantum system with the potential sequence  $v_n$.
The property  corresponds to the change of the potential sequence
from the non-Gaussian stationary process 
to the nonstationary process at $B=2$.
As to cover the anomalous fluctuation of the classical system by the quantum 
effect, the result of the quantum system is that the Lyapunov exponent 
is still positive (i.e. the localization length of the initially localized 
wavepacket is finite). 
In addition, the NLL  $\Lambda_N$ is shown in Fig.\ref{fig:E0-B-dep}(b)  
in order to investigate the details of the size effect on the localized states 
for $B \geqq 2 $. 
Accordingly,  as we can infer from the behaviour of $\gamma_N$, 
the NLL also has an exponential dependence on $B$.


\begin{figure}[htbp]
\begin{center}
\includegraphics[width=7.0cm]{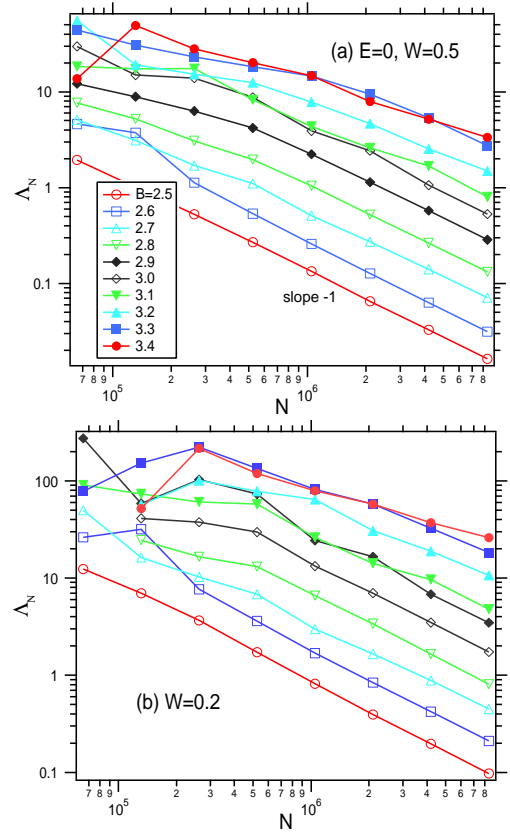}
\caption{(Color online)
The normalized localization length $\Lambda_N$ as a function of 
the system size $N$ for 
(a)$W=0.5$, (b) $W=0.2$.
The other parameters are the same as Fig.\ref{fig:E0-B-dep}.
Note that all axes are in logarithmic scale.
}
\label{fig:E0-N-dep}
\end{center}
\end{figure}

\begin{figure}[htbp]
\begin{center}
\includegraphics[width=5.0cm]{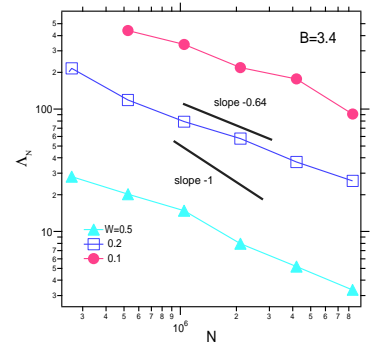}
\caption{(Color online)
The normalized localization length $\Lambda_N$ as a function of 
the system size $N$ for $B=3.4$ with $W=0.1,0.2, 0.5$.
Note that both axes are in logarithmic scale.
The lines with slope $-1$ and $-0.64$ are shown as a reference.
}
\label{fig:E0-N-dep-2}
\end{center}
\end{figure}

\subsection{
Power-law localized behaviour
}
Figure \ref{fig:E0-N-dep} shows the $N-$dependence of NLL.
In the case of  $B=2.5$, it obviously declines with $N^{-1}$ irrespective of 
$W=0.5,0.2$.
This corresponds to the clearly exponential localization at this system size, 
as could be seen from the definite value of  $\gamma_N >0$.
On the other hand, as $B$ increases, 
a tendency to decrease with the exponent $0<\eta <1 $ can be seen as
\beq
  \Lambda_N \sim N^{-\eta}.
\eeq
Figure \ref{fig:E0-N-dep-2} shows $N-$dependence of  $\Lambda_N$
for some $W$ with $B=3.4$.
Estimating these slopes by the least-squares method shows that $\eta \simeq 0.64$
 is almost independent of the value of $W$.
Although this behaviour remains present until the infinity of $N$,  
the wavefunction might still  be normalized in this region,
 and behave like power-law localized states 
with the point spectrum since $\delta > 0.5$.
However, there is a possibility that exists
a characteristic length separating the outer exponential decay 
from the inner power-law decay of the localized states
as in  two-dimensional disordered systems
with large localization length. 
That is, the wavefunction is power-law decay inside the localization length and 
exponential decay in the outside.
It should be noticed as well, that the power-law behaviour with $ 0 <\eta <1$ is observed 
as a transient phenomenon asymptotically going to $N^{-1}$ ($\eta=1$)
for $N \to \infty$. 
Detailed calculations with larger system size are required for the definite conclusions.




\section{Quantum wavepacket dynamics}
\label{sec:DL}
This is just the main section of the present paper.
LDT has been observed in phase space $(\alpha, W)$ and $(D, W)$ 
based on the Lyapunov exponent and/or NLL 
 with FFM potential and Weierstrass potential, respectively.
However, the effect of the long-range correlation on quantum diffusion
in the 1DDS have not sufficiently been studied yet in 1DDS with long-range correlation.
In this section, we examine the dynamical property  
of the initially localized wave packet by changing the 
parameter $B$ and $W$, but how would the SNCT characterized by the $1/f$ fluctuation
influence on the localization property of the quantum states?
Generally speaking,  the  positive Lyapunov exponent value does not seem to be the sole 
 sufficient condition for the dynamical localization to occur.
Hence, it turns out that directly studying the wavepacket dynamics is very important.

\subsection{Method}
The quantum time-evolution is given by  
\beq
i\hbar \frac{\pr \phi(n,t)}{\pr t}= \phi(n+1,t)+\phi(n-1,t) + Wv(n)\phi(n,t),
\eeq
where $ n=1,2,...,N$.
We take the initial state to be localized at the centre of the system, 
$\phi(n,t=0)=\delta_{n,N/2}$,  
and $\hbar=1$ throughout this calculation.
In the numerical calculation we used FFT-symplectic integrator (SI) scheme
for integrating the time-dependent Schrödinger equation \cite{takahashi97,yamada99}.
Then, we adapted periodic boundary condition because
we used the FFT and inverse FFT to exchange the real space representation
for the momentum space one and do the opposite operation.
However, here we only consider the time periods,  when the boundary does not influence
the essential results.

We characterize the spread of the wavepacket by 
the mean square displacement (MSD),
\begin{eqnarray}
m_2(t) = \sum_{n}(n-n_0)^2 \left< |\phi(n,t)|^2 \right>.
\end{eqnarray}
In general, in the long-time limit the time-dependence is given as 
\beq
   m_2 \sim t^\sigma,
\eeq
where $\sigma$ is diffusion exponent.
$\sigma=0$ corresponds to the localization, 
$0< \sigma <1$ to subdiffusion,
$\sigma =1$ to normal diffusion,
$1< \sigma <2$ to superdiffusion,
and $\sigma =2$ to ballistic motion.
Moreover,  $\sigma=2/3$ should correspond to the metal-insulator transition in 3DDS \cite{vollhardt92},
whereas $\sigma =1$ and $\sigma =2$ might also be found in 1DDS 
with the stochastically fluctuation and in the periodic systems, respectively.
Furthermore, superdiffusive motions ($1<\sigma<2$) can be observed in 
quantum chaotic systems as well. 


In appendix \ref{app:average}, we added more detailed data for several cases
in order to confirm the numerical  stability of the ensemble average computation for the MSD.

\begin{figure}[htbp]
\begin{center}
\includegraphics[width=8cm]{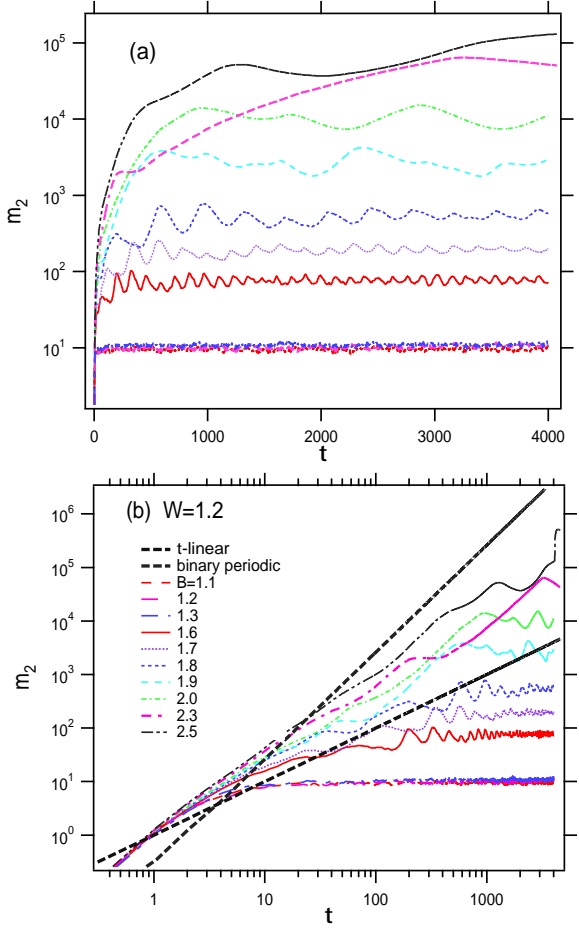}
\caption{
(Color online)
(a)The second moment $m_2$ as a function of time for some values of 
$B$ at $W=1.2$.
Note that the vertical axis is in the logarithmic scale. 
(b)The double-logarithmic plot.
The ballistic motion ($m_2 \sim t^2$) of the binary periodic case and
diffusive lines ($m_2 \sim t^1$) are shown as a reference.
The results for $B=1.1,1.2,1.3$ in the bottom are almost overlapped.
We set $\hbar=1$ and $\delta t=0.05$ through this paper.
The system size and sample size are $N=2^{14}$ and $100$, respectively, 
in the numerical calculation in this section.
}
\label{fig:msd-B-1}
\end{center}
\end{figure}

\begin{figure}[htbp]
\begin{center}
\includegraphics[width=8cm]{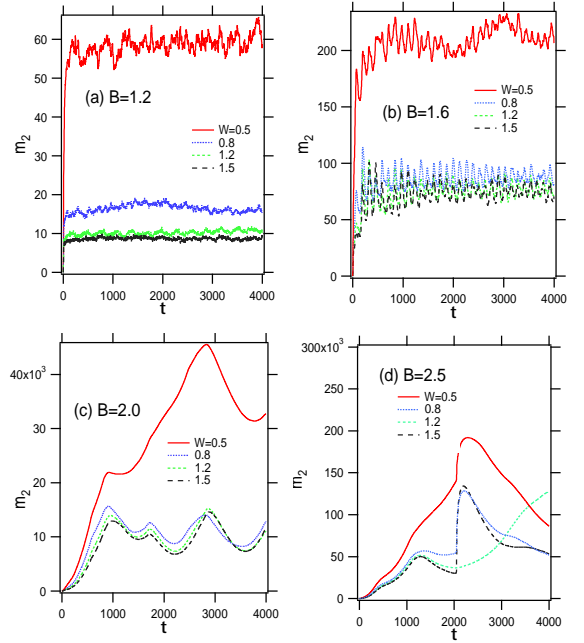}
\caption{
(Color online)
The second moment $m_2$ as a function of time for some values of 
$W$ at (a)$B=1.2$, (b)$B=1.6$, (c)$B=2.0$ and (d)$B=2.5$.
Note that the both axes are in the real scale. 
}
\label{fig:msd-W}
\end{center}
\end{figure}

\subsection{Stationary regime ($1<B \leq 2$)}
Figure \ref{fig:msd-B-1} shows the time-dependence of MSD 
 for the potential strength $W=1.2$ 
 with several values of the correlation parameter $B$.
First consider only the stationary regime of $B <2$.
We may note that  after long time the system evolves 
from the ballistic-like rise ($m_2 \sim t^2$) 
to the complete localization  ($m_2 \sim t^0$),  
see the double-logarithmic plot in Fig.\ref{fig:msd-B-1}(b).
Furthermore, in the cases of $B=1.1,1.2, 1.3$, there is no distinction 
in the wavepacket spreading degree.
This can be easily understood, since the power spectra of the system 
are white-noise regardless of  $B$ in the regime ($1<B<3/2$).
Meanwhile,  the wavepacket spreading degree increases 
along with the $B$ increase in the regime of $ 3/2 <B <2 $.


On the other hand, Fig.\ref{fig:msd-W}(a)-(c) show the resulting time 
dependence of MSD with changing $W$.
In the white noise regime, spread of the wavepacket decreases when $W$ increases, 
but in the region of $ 3/2 <B <2 $, 
the larger the value of $B$, the larger fluctuation amplitude of 
the MSD time dependence.
This gets much more apparent when $W$ is decreasing.
Furthermore, even in $B=2$ case, that is,  in the nonstationary regime,
the localization would seem to result from the ballistic rise, as shown in Fig.\ref{fig:msd-W}(c).




\subsection{Nonstationary regime ($B \geqq 2$)}
\begin{figure}[htbp]
\begin{center}
\includegraphics[width=7cm]{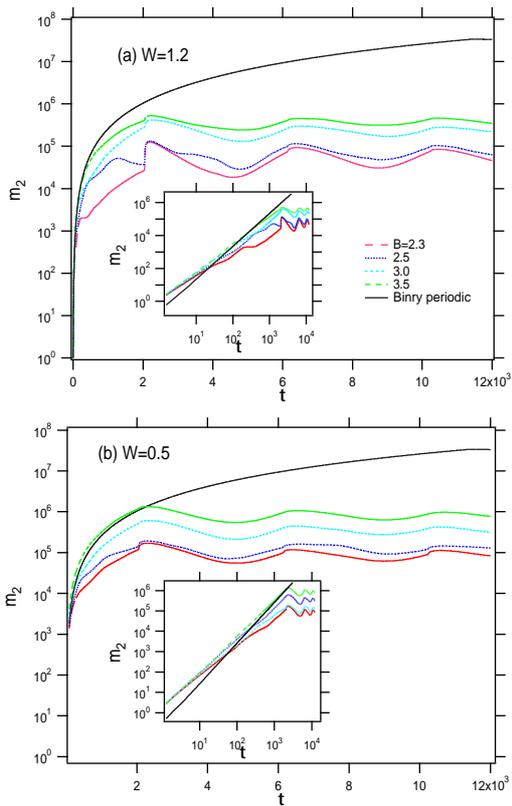}
\caption{
(Color online)
The second moment $m_2$ as a function of time for some values of 
$B$ at (a)$W=1.2$ and (b)$W=0.5$.
The ballistic increase ($m_2 \sim t^2$) of the binary periodic system is also plotted by 
black bold line as a reference.
Note that the vertical axis is in the logarithmic scale, and 
both  axes of the inset in the panels (a) and (b) are in the logarithmic scale. 
}
\label{fig:long-balli-2}
\end{center}
\end{figure}

Consider the result of quantum diffusion in the nonstationary regime.
As shown in Fig.\ref{fig:msd-B-1}(b) and Fig.\ref{fig:msd-W}(d), 
it is complicated and its behaviour is unclear, 
and the results of long-time calculations with the fixed value of $W =1.2$
are shown in Fig.\ref{fig:long-balli-2}(a).
There, the results in the case of a binary periodic sequence ($v_n=(-1)^n$) 
showing complete ballistic motion ( $m_2 \propto t^2$) are also shown as a reference.
It seems that the MSD  increases with time in the ballistic regime and 
 localizes after passing through the intermediate spreading regime.
(See the insets in Fig.\ref{fig:long-balli-2}(a) and (b).)
The spread is larger along with the rise in $B$, and the fluctuation 
becomes longer than that found in stationary regime  $B <2$.
%
Furthermore, the result in the case of the potential strength $W =0.5$ 
is shown in Fig.\ref{fig:long-balli-2}(b).
It can be noted that it dynamically localizes at $t \to \infty$, 
even if it is in nonstationary regime regardless of $W$.
As a result, dynamical localization also occurs 
in these regimen investigated, so that it is in full agreement with our findings for 
the Lyapunov exponents and NLL shown in the previous section.

\subsection{
Scaling of the localization dynamics
}
\begin{figure}[htbp]
\begin{center}
\includegraphics[width=8cm]{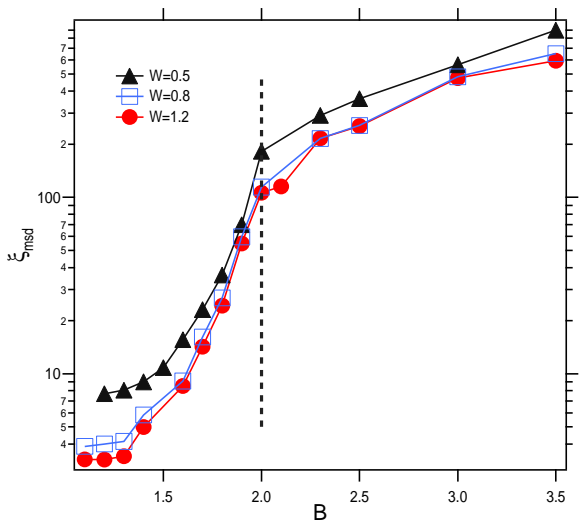}
\caption{
(Color online)
Dynamical localization length $\xi_{msd}$ as a function of the correlation parameter $B$
for $W=0.5,0.8,1.2$, which are determined by ensemble-average and time-average for a period 
with the stable fluctuation.
The $B-$dependence changes around $B \simeq 2$  from rapid increase to milder one.
}
\label{fig:LL}
\end{center}
\end{figure}

The localization length becomes definite only for the perfect localization ($\sigma=0$),
and infinity for $\sigma>0$,  in a limit $t \to \infty$.
Here, we define the DLL $\xi_{msd}$ by MSD as 
$m_2(t \to \infty)=\xi_{msd}^2$.
Figure \ref{fig:LL} shows the $B-$dependence of 
the DLL $\xi_{msd}$ numerically obtained as a time-averaged value 
from the data that has entered the fluctuation for a long time.
It turns out to be increased sharply towards $B=2$.
%
Even in nonstationary regime $B \geqq 2$, 
the localization length also increases along with the increase in $B$, 
but its $B$-dependence is being changed more moderately when $B >2$.
Regardless of the value of $W$, the similar change in the DLL 
around  $B \simeq 2$ can be seen in Fig.\ref {fig:LL}.
How does the change in the correlation of potential 
sequence affect the localization dynamics of the quantum system
with the potential near the SNCT?
The numerical results suggest that   
it does not change the qualitative nature of the quantum states, 
but we can nonetheless state that the parameter dependence 
experiences a change in this case.


As a result, in all the regions in question including the nonstationary regime 
 the rise of the initially localized wavepacket is ballistic of $m_2 \sim t^2$,
and it localized at finite DLL $\xi_{msd}$ 
by passing through an increasing region 
with an index close to 2 but less than 2 ($\sigma \leq 2$),
as seen in Fig.\ref{fig:long-balli-2}.
In this case, instead of the MSD, we introduce 
the scaled MSD
\begin{eqnarray}
\Lambda(t)\equiv\frac{m_{2}(t)}{t^2}.
\label{eq:scale-1}
\end{eqnarray}
This type MSD was introduced to investigate LDT phenomena 
at the critical point \cite{delande07}.
$\Lambda =const$ if the wavepacket is extended, i.e. $m_2 \sim t^2$.
If this localization dynamics can be scaled only by the DLL  $\xi_{msd}$, 
the scaling for the $t-$dependence of the scaled MSD can be expected to be as follows:
\beq
  \Lambda(t,B,W) = F\left( \frac{t}{\xi_{msd}(B,W)} \right), 
\label{eq:scaling-1}
\eeq
where the asymptotic form of the scaling function is 
\beq
F(x) \sim       
  \begin{cases}
    const     & x \to 0\\
    \frac{1}{x^2}  & x \to \infty.
  \end{cases}
\label{eq:scaling-2}
\eeq

\begin{figure}[htbp]
\begin{center}
\includegraphics[width=7cm]{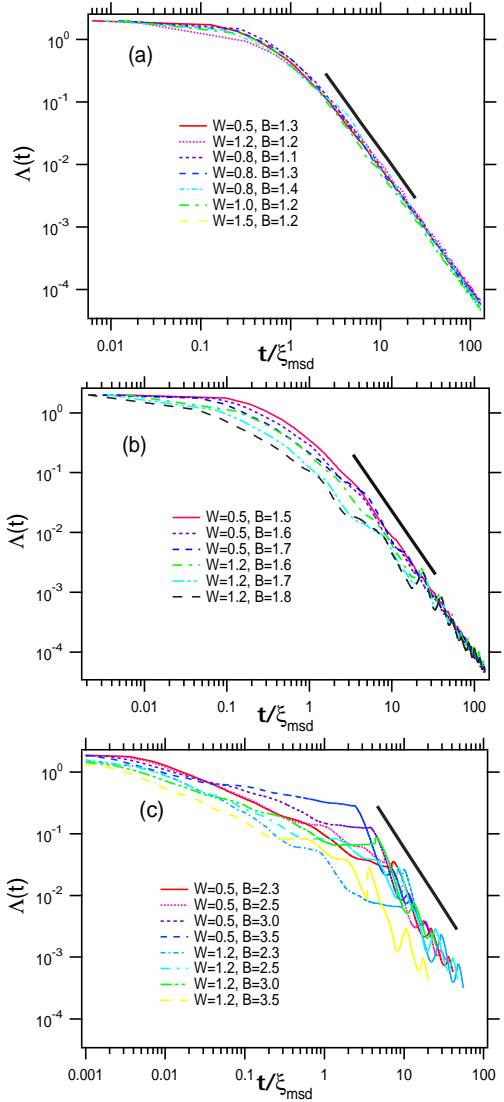}
\caption{
(Color online)
Scaled MSD $\Lambda(t)$ as a function of 
the scaled time $t/\xi_{msd}(B,W)$
by the localization length $\xi_{msd}$.
(a)White noise regime ($B<3/2$).
(b)non-Gaussian stationary regime ($3/2<B<2$).
(c)nonstationary regime ($B>2$).
The lines with slope -2 are shown as a reference.
Note that all the axes are in the logarithmic scale.
}
\label{fig:mb-scale-1}
\end{center}
\end{figure}

If the one-parameter scaling is true for the localization phenomena of the wavepacket
the scaled MSD smoothly connects the asymptotic behavior in Eq.(\ref{eq:scaling-2}).
Figure \ref{fig:mb-scale-1} shows $\Lambda(t)$ as a function of 
the scaled time $t/\xi(\alpha)$ to characterize the  localization phenomenon
at various $W$ and $B$ for each of the three regions.
Hence, it is then clear that the case of  $B<3/2$ is bridging the space from the ballistic 
region to the localized region even in the various combinations of  $(W, B)$.
These results support that the localization process 
from the ballistic motion ($\sim t^2$) to the localization ($\sim t^0$) in MSD
of the wavepacket  is scaled 
by a single parameter, i.e. by DLL $\xi_{msd}$. 

On the other hand, as seen in the Fig.\ref{fig:mb-scale-1}(b), 
 in the nonGaussian stationary regime ($3/2<B<2$), 
it is found that in the transient region from the ballistic motion
 region to the region of the localization the magnitude of the deviation 
becomes large although  the one parameter scaling is true 
for sufficiently long times when localisation becomes apparent.
Figure \ref{fig:mb-scale-1}(c) shows the result for the non-stationary region.
The deviation in this transient region becomes larger.



\section{Summary and discussion}
\label{sect:summary}
In summary, we numerically studied the nature of  localized property  
 in 1DDS with long-range correlation generated by modified Bernoulli map,
paying attention to around $B = 2$ where anomalous fluctuation of $1/f$
 occurs in the potential sequence.
First, we used Lyapunov exponent and the normalized localization length to
investigate the localized behaviour of the wavefunctions for the band centre energy $E=0$.
Lyapunov exponent linearly decreases as $B$ approaches 2 for $3/2<B<2$, 
and changes from the linear-decay to the exponential-decay for $B>2$.
Next, we have investigated dynamical property of the initially localized wavepacket.
As a result, we find that the wavepacket localizes irrespective of the stationary ($3/2<B<2$)
and nonstationary regimen ($B \geqq 2$).
However, it was shown that the $B-$dependence of the dynamical 
localization length determined by 
MSD changes from rapid growth to the slower one around the SNCT $B \simeq 2$.
Moreover we show that the localization dynamics for $B>3/2$ deviates from the 
one-parameter scaling theory of the localization in the transient region from ballistic
to localization.. 
Analysis on the dynamics of the localization is still very few in 
other 1DDS with long-range correlation \cite{santos06}.

The basic property of the quantum states is directly related to 
physical phenomena such as electronic conduction and transmission.
In practice, we observed the physical quantities such as electronic transmission and
acoustic wave localization in one-dimensional systems 
with sequence generated by chaotic maps \cite{esmailpour08,moura11,diaz05,shahbaz05}.
Also,  real DNA chains can be described by the four symbolized 
sequence as "A", "C","G","T, ", and 
both coding and/or non-coding DNA sequences can be treated 
as those having the long-range correlation
with the power spectrum $S(f) \sim 1/f^\alpha$($0.8 <\alpha<1.2$) 
for lower $f$ \cite{dna92,dna02,dna03,guo07,dna08,dna09,yamada04a}.
Accordingly, it is expected that 
the localization-delocalization  problem is strongly related to 
electronic conduction in the DNA chains.
We expect that the present work stimulate further studies of delocalized states and 
the localization-delocalization transition in correlated disordered systems.

\appendix

\section{Ensemble average}
\label{app:average}

\begin{figure}[htbp]
\begin{center}
\includegraphics[width=7cm]{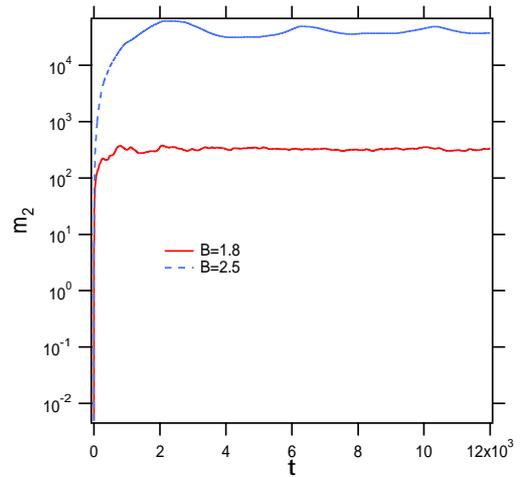}
\caption{
(Color online)
MSD as a function of time for $W=1.2$ with $B=1.8, 2.5$.
The average is taken over 1000 different initial values of the map.
}
\label{fig:sample}
\end{center}
\end{figure}

Figure \ref{fig:sample} shows time-dependence of MSD
for  $B=1.8$ and $B=2.5$, which are taken ensemble average
over 1000 different initial condition of the map.
The global behaviour is the same as the case in text for
 $B=1.8$ and $B=2.5$.
The fluctuation remains after the average and 
it is larger for the larger value of $B$.
In the text, 
the dynamical localization length $\xi_{msd}$ based on the time-dependence 
is determined 
by taking the time average for the data in the fluctuation regime.


\section*{Acknowledgments}
The author would like to thank Professor M. Goda for discussion 
about the correlation-induced delocalization at 
early stage of this study, 
and  Professor E.B. Starikov for proof reading of the manuscript.
The author also would like to acknowledge the hospitality of 
the Physics Division of The Nippon Dental University at Niigata
for  my stay, where part of this work was completed.

\section*{Author contribution statement}
The sole author had responsibility for all parts of the manuscript.



\end{document}